\documentclass[aip, preprint]{revtex4-1}
\usepackage{amsbsy,amssymb,amsmath,bm, gensymb}
\usepackage{graphicx,color,epsfig,rotate}

\usepackage{setspace}

\begin{document}

\topmargin -15mm

\title{Effect of stoichiometric vacancies on the structure and properties of the Ga$_2$SeTe$_2$ compound semiconductor}


\author{N. M. Abdul-Jabbar}
\affiliation{Department of Nuclear Engineering, University of California, Berkeley, California 94720, USA}
\affiliation{Materials Sciences Division, Lawrence Berkeley National Laboratory, Berkeley, California 94720, USA}

\author{T. R. Forrest}
\affiliation{European Synchrotron Radiation Facility, BP 220, F-38043 Grenoble Cedex, France}
\affiliation{Department of Physics, University of California, Berkeley, California 94720, USA}

\author{R. Gronsky}
\affiliation{Department of Materials Science and Engineering, University of California, Berkeley, California, 94720, USA}

\author{E. D. Bourret-Courchesne}
\affiliation{Materials Sciences Division, Lawrence Berkeley National Laboratory, Berkeley, California 94720, USA}

\author{B. D. Wirth}
\affiliation{Department of Nuclear Engineering, University of California, Berkeley, California 94720, USA}
\affiliation{Department of Nuclear Engineering, University of Tennessee, Knoxville, Tennessee 37996, USA}

\begin{abstract}

Ga$_2$SeTe$_2$ belongs to a family of materials with large intrinsic vacancy concentrations that are being actively studied due to their potential for diverse applications that include thermoelectrics and phase-change memory. In this article, the Ga$_2$SeTe$_2$ structure is investigated via synchrotron x-ray diffraction, electron microscopy, and x-ray absorption experiments. Diffraction and microscopy measurements showed that the extent of vacancy ordering in  Ga$_2$SeTe$_2$ is highly dependent on thermal annealing. It is posited that stoichiometric vacancies play a role in local atomic distortions in Ga$_2$SeTe$_2$ (based on the fine structure signals in the collected x-ray absorption spectra). The effect of vacancy ordering on Ga$_2$SeTe$_2$ material properties is also examined through band gap and Hall effect measurements, which reveal that the Ga$_2$SeTe$_2$ band gap redshifts by $\approx$0.05 eV as the vacancies order and accompanied by gains in charge carrier mobility. The results serve as an encouraging example of altering material properties via intrinsic structural rearrangement as opposed to extrinsic means such as doping. 

\end{abstract}

\maketitle

\section{Introduction}

Ga$_2$SeTe$_2$ is a compound semiconductor that belongs to a class of III-VI materials that exhibit a cubic zincblende crystal structure (F\={4}3\emph{m} space group) dominated by stoichiometric or structural vacancies (otherwise known as defect zincblende). These defects arise due to the valence mismatch between the anion and cation sites forcing 1/3 of the cation sites to be vacant in order to electrically stabilize the crystal. Recent investigations on binary III-VI materials with defect zincblende structures have shown that their stoichiometric vacancies can lead to material properties that are suitable for a vast number of applications. Examples include observing ordered vacancy planes in Ga$_2$Te$_3$ that serve as effective phonon scatterers that decrease thermal conductivity, which is crucial for thermoelectric applications \cite{Kurosaki:2008cj, Kim:2009dz, Yamanaka:2009ka}. Additional work on Ga$_2$Te$_3$ has also shown it could be an attractive material for phase-change random access memory applications as it shows better data retention ability, low power consumption, and high dynamical electric switching ratios when compared to the more widely studied Ge$_2$Sb$_2$Te$_5$ \cite{Zhu:2010jj}. The presence of stoichiometric vacancies in this class of material also leads to a loose crystal structure, which allows for anomalously high radiation stability by minimizing Frenkel pair production from incident radiation; parameters such as charge carrier concentration, charge carrier mobility, and microhardness measured before and after irradiation showed little or no change \cite{Koshkin:1994, Koshkin:vp, Koshkin:1976wj}. As a result, such materials may also have potential as nuclear particle detectors for high energy physics or security applications.

The ternary Ga$_2$SeTe$_2$ compound is expected to display properties analogous to III-VI binary zincblende compounds, but with the added benefit of band gap engineering \cite{Huang:2014fo}; though one must be aware of the solid state immiscibility in the Se rich region of the Ga$_2$Te$_3$-Ga$_2$Se$_3$ phase diagram \cite{Warren:1974ux}. Nevertheless, there is a paucity of information on the Ga$_2$SeTe$_2$ compound semiconductor; here, we present an experimental investigation to study the influence of stoichiometric vacancies on the structure and physical properties of Ga$_2$SeTe$_2$. We utilize high-resolution single crystal diffraction experiments and transmission electron microscopy (TEM) to characterize the basic crystal structure of Ga$_2$SeTe$_2$. Then we attempt to probe the effect of vacancies on local structure via extended x-ray absorption fine structure (EXAFS) experiments. Finally, we examine the effect of stoichiometric vacancies on material properties by correlating our structural results with band gap and charge carrier mobility measurements.

\section{Experimental Methods}

Ga$_2$SeTe$_2$ single crystals were grown using a modified Bridgman method from stoichiometric amounts of 8N Ga, 6N Se, and 6N Te based on a procedure we reported elsewhere \cite{AbdulJabbar:2012bma}. Single crystalline specimens harvested from the solidified ingot were subjected to three thermal treatments (similar treatments have been observed to alter the vacancy distributions in powder Ga$_2$Te$_3$ specimens \cite{Kim:2011kb}): Single crystalline and powder specimens harvested from the solidified ingot and were subject to three thermal treatments: (1) $735\,^{\circ}\mathrm{C}$ anneal for 28 days followed by quenching to $0\,^{\circ}\mathrm{C}$, (2) $435\,^{\circ}\mathrm{C}$ for 28 days followed by slow cooling in the furnace, and (3) sample in the as-grown state.

Single crystal x-ray diffraction experiments were performed at room-temperature on the bending magnet beam line 33-BM at the Advanced Photon Source (APS) at Argonne National Laboratory (ANL) employing a focused beam of 15 keV x rays. Single crystal Ga$_2$SeTe$_2$ specimens were mounted on a 4-circle diffractometer in an off-specular geometry utilizing a scintillation area detector for data collection. Detailed information on the instrument setup has been reported elsewhere \cite{Karapetrova:2011tb}.

Electron microscopy was carried at the National Center for Electron Microscopy (NCEM) at Lawrence Berkeley National Laboratory. Images were collected using a modified Philips CM300FEG/UT electron microscope at an operating voltage of 300 kV. Sample preparation involved dispersing crushed Ga$_2$SeTe$_2$ single crystals in an isopropanol solution on conventional Au TEM grids. 

Extended x-ray absorption fine structure (EXAFS) experiments were performed at the 05-BM bending magnet beam line at the APS and the 4-1 bending magnet beam line at the Stanford Synchrotron Radiation Lightsource (SSRL) at SLAC National Accelerator Laboratory. Powders ground from single crystal Ga$_2$SeTe$_2$ specimens (measured in the single diffraction experiments) and sieved to 60 $\mu$m were dispersed onto scotch tape and mounted onto the x-ray beam in a transmission geometry. Scans were taken by tuning the beam to the Ga (10.367 keV), Se (12.658 keV), and Te (31.814 keV) K-edges (five scans were taken at each x-ray energy to minimize statistical errors). Data analysis was done via the SIXpack software package \cite{Webb:2005jv}.

The band gaps of Ga$_2$SeTe$_2$ samples were measured using a Perkin-Elmer 950 Lambda spectrophotometer. Transmission and reflectance data from the polished Ga$_2$SeTe$_2$ single crystal samples were used to plot optical absorption edges, where the rise in absorption was used to determine the band gap. 

Electrical properties of Ga$_2$SeTe$_2$ were measured via Hall effect using Ga$_2$SeTe$_2$ single crystals with surface area of $\approx$ 4 mm$^2$ and thickness of $\approx$1 mm. A four-probe van der Pauw geometry (utilizing contacts made from silver paste) was implemented. Measurements were done under a field of 1.0 T using a 1 nA excitation current.

\section{Results and Discussion}

X-ray diffraction experiments on as-grown, 435 \degree C, and 735 \degree C annealed Ga$_2$SeTe$_2$ crystals show a cubic zincblende structure with a lattice constant of $a$ = 5.77 \r{A}. However, high-resolution reciprocal lattice mapping (shown in Figure 1) reveals that specimen thermal history has a profound effect on the secondary structural characteristics of Ga$_2$SeTe$_2$. Mainly, in the 735 \degree C annealed specimen, well-defined satellite reflections are observed around the fundamental Bragg reflections. These occur in pairs at 1/16[2 1 0] in equivalent directions from the centrally located Bragg peak. These ancillary reflections originate from superstructures associated with the stoichiometric vacancies in the defect zincblende crystal system. Their locations in reciprocal space suggest long-rage modulated structure parallel to $<$210$>$. However, dark field TEM imaging (see Figure 2) shows only the presence of two-dimensional vacancy structures parallel to $<$111$>$ directions that order at $\approx$2.7 nm intervals or about 8 lattice units (we have reported results from high-resolution TEM investigation on  Ga$_2$SeTe$_2$ elsewhere \cite{AbdulJabbar:2014fl}). Such a discrepancy has been previously observed in as-grown Ga$_2$Te$_3$ \cite{Otaki:2009if, Otaki:2009cg, Kashida:2009bc}. Hence, the x-ray data suggest that the stoichiometric vacancy structures in Ga$_2$SeTe$_2$ are based on coupled displacive (as directly observed via TEM) and atomic modulations. 

Inspecting the reciprocal lattice maps for the as-grown and 435 \degree C annealed Ga$_2$SeTe$_2$ crystals, we see an absence of distinct satellite reflections. Here in the as-grown condition, the two-dimensional vacancy structures in these crystals lose their long-range spatial ordering. Nevertheless, they retain their proclivity to diffract x rays evidenced by the diffuse scattering features present around the main Bragg reflections, which is indicative of short-range periodicity.

\begin{figure*}[ht]
\includegraphics[width=\textwidth]{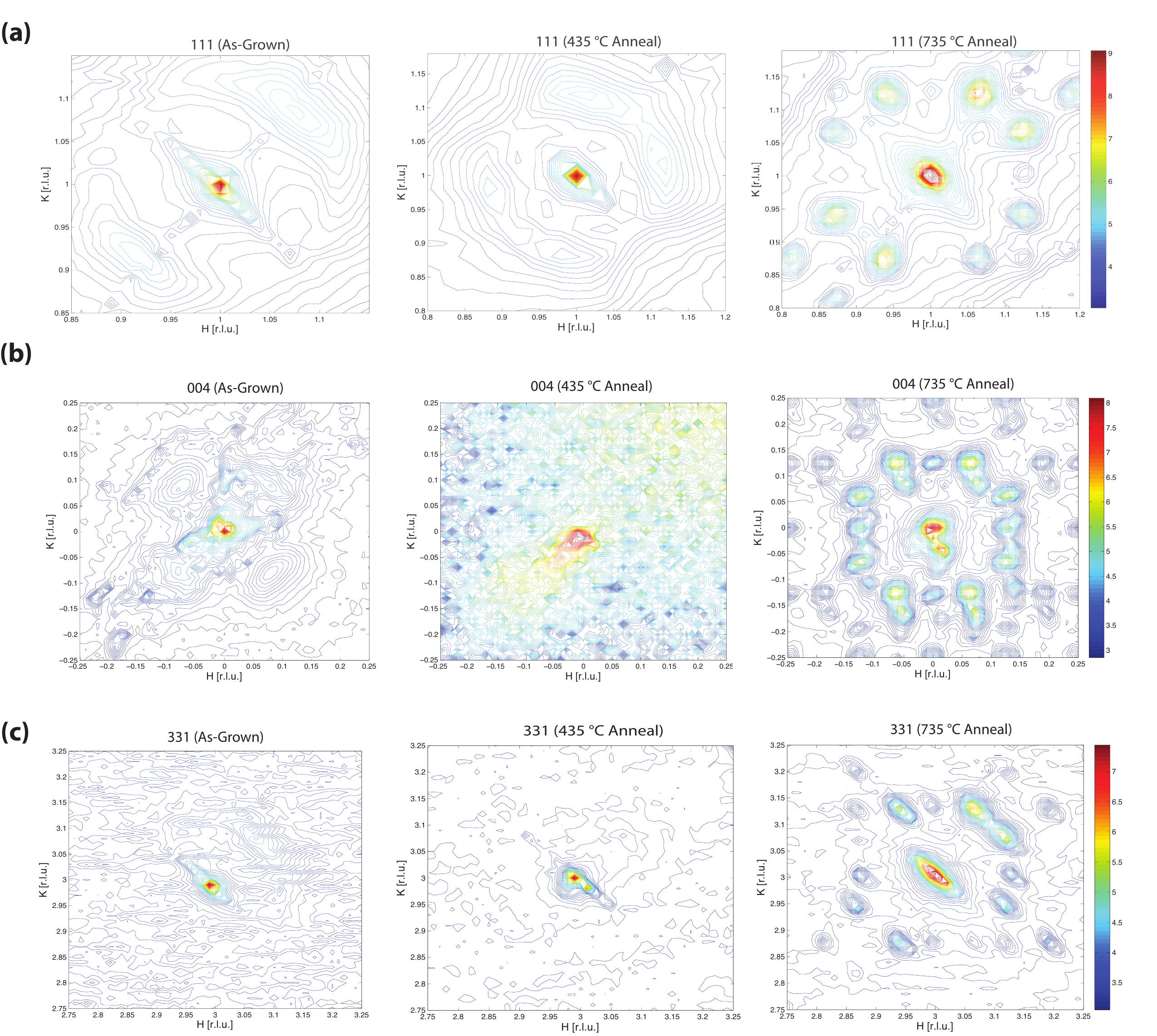}
\caption{Reciprocal lattice scans at the 111 (a), 004 (b), and 331 (c) Bragg reflections for as-grown, 435 \degree C, and 735 \degree C annealed Ga$_2$SeTe$_2$ single crystals}
\label{fig1}
\end{figure*}

\begin{figure}
\includegraphics[width=\textwidth]{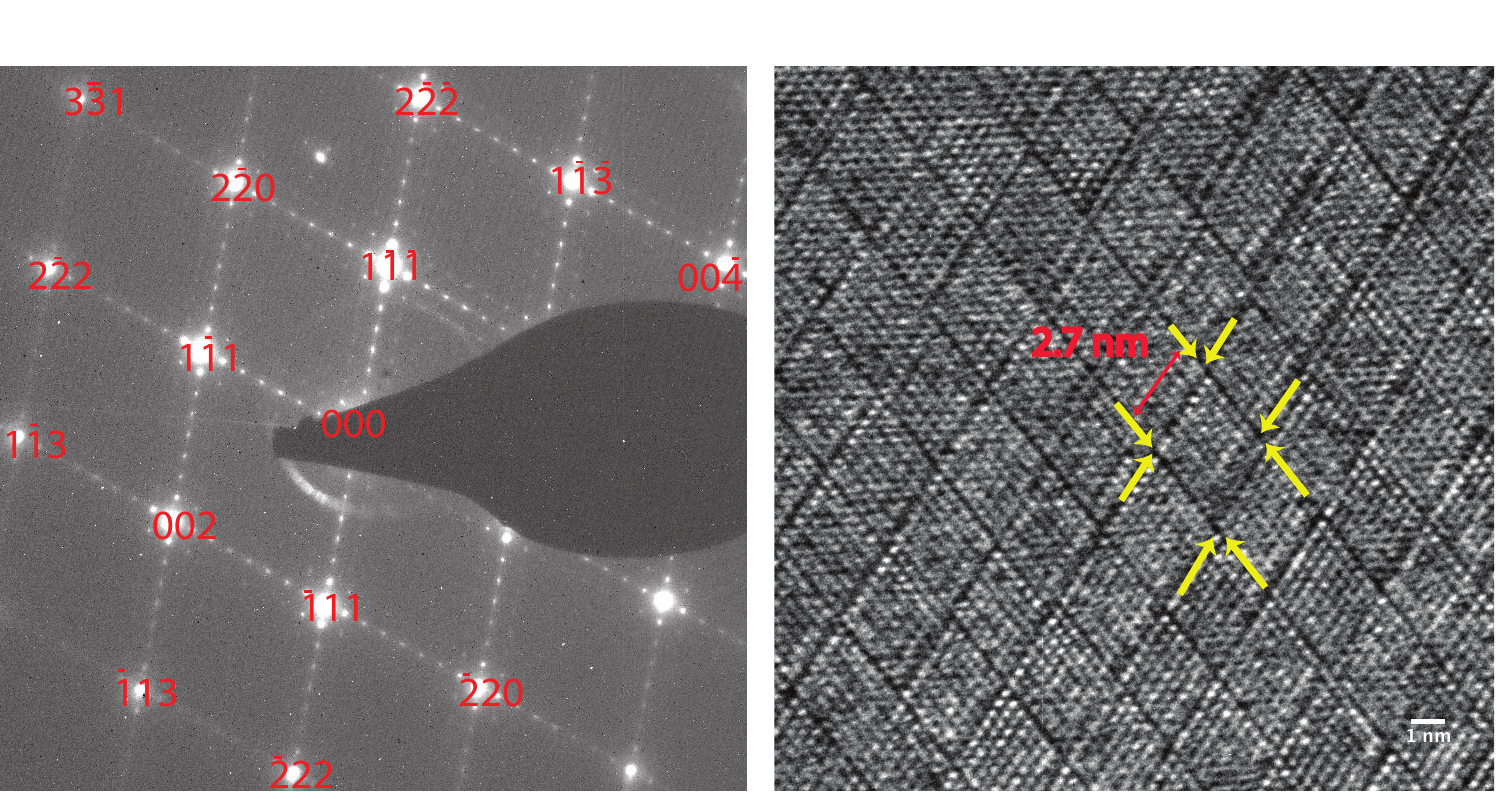}
\caption{Electron diffraction pattern and dark field TEM image of 735 \degree C annealed Ga$_2$SeTe$_2$ in the [110] zone axis. Ordered two-dimensional vacancy structures are represented by a periodic dark line contrast that propagates in $<$111$>$ directions at 2.7 nm intervals. This correlates with the electron diffraction pattern, where the satellite reflections occur at 1/8 of the lattice spacing in reciprocal space.}
\label{fig2}
\end{figure}

A common feature across all samples in the reciprocal lattice maps is the streaking of the Bragg peaks diagonally in $h$-$k$ space, alluding to crystal distortions arising from the stoichiometric-based vacancy structures; though global symmetry of the Bravais lattice associated with a zincblende crystal is preserved. This is captured in the reciprocal lattice maps and TEM micrographs shown in Figures 1-2. As a result, we believe that the elongation of the Bragg peaks arises from local symmetry distortions. To confirm this, we performed extended x-ray absorption fine structure measurements to probe the local structure of Ga$_2$SeTe$_2$. This is done by tuning an incident x-ray beam to the K-edge energies ($i$.$e$. energy required to eject the 1$s$ electrons from the K shell) of the constituent atoms of the specimen; when x rays are absorbed by the atom, a photoelectron is emitted and propagates in the lattice as a spherical wave where interference can occur between outgoing and backscattered waves \cite{AlsNielsen:2011vn}. This phenomenon gives rise to oscillations in the observed absorption cross-section, which yields information on the local structure of the absorbing atom. Taking the Fourier transform of the fine structure oscillations in the absorption spectra yields radial distribution functions with peaks corresponding to the nearest neighbor atomic shells around the absorbing atom. These are shown in Figure 3 at the Te, Se, and Ga K-edge energies, which correspond to the local structure around Te, Se, and Ga atoms respectively.

Since Ga$_2$SeTe$_2$ is formed via Ga-Te and Ga-Se tetrahedra packed in the zincblende structure, we expect the nearest neighbor shells around the Ga, Se, and Te atoms to correspond to the cation-anion (Ga-Se or Ga-Te) bond distance, which is 2.498 \r{A}. Inspecting the Te K-edge for the as-grown and annealed samples, however, the dominant scattering shell occurs at $\approx$2.27 \r{A}, indicative of a contraction of the Ga-Te atomic dumbbell. Similarly at the Se K-edge, the common peak is observed at $\approx$2.14 \r{A} representing the Ga-Se bond length. Compared with Ga-Te bond length, a larger contraction is likely because the Se atom is lighter. Additionally, real scattering shells are observed at $\approx$1.80 \r{A} for all Ga$_2$SeTe$_2$ samples not associated with the cubic lattice at the Te and Se K-edge energies. Local structural effects are notably prominent as we inspect the radial distribution functions for the Ga K-edge; this is not surprising as the stoichiometric vacancies in Ga$_2$SeTe$_2$  originate at the Ga site. Here the common peak corresponding to the cation-anion dumbbell occurs at $\approx$2.43 \r{A}, once again suggesting local atomic contraction. Unlike the local environment around Se or Te, however, the dominant local structural features predominantly occur in the range of 1-2 \r{A}--providing additional evidence of the local lattice distortions likely caused by stoichiometric vacancies. The exact mechanism driving such distortions and the role of vacancy periodicity remains uncertain. Earlier studies on Ga$_2$Te$_3$ have posited that stoichometric Ga vacancies can induce Jahn-Teller lattice distortions that drive a transition to a tetragonal structure \cite{Otaki:2009if, Otaki:2009cg, Kashida:2009bc}. We were not able to verify this for Ga$_2$SeTe$_2$, but based on our x-ray absorption data, we confirm that distortions caused by stoichiometric vacancies are real and occur locally within the crystal. 

\begin{figure*}
\includegraphics[width=\textwidth]{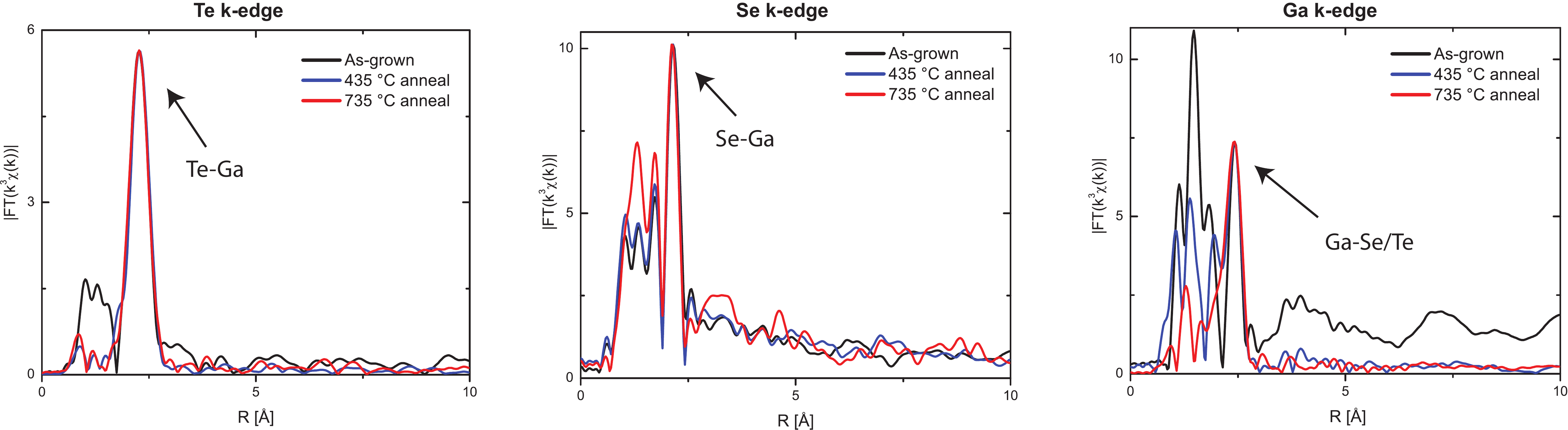}
\caption{ Radial distribution functions obtained from EXAFS oscillations that show the local structure around the absorbing Te, Se, and Ga atoms. The common peak corresponding to nearest neighbor Ga-Te and Ga-Se bonds in the zincblende crystal are arrowed.}
\label{fig3}
\end{figure*}

Our x-ray measurements show that thermal history plays a prominent role in the distribution of stoichiometric vacancies in Ga$_2$SeTe$_2$. Yet information on how the distribution of stoichiometric vacancies affect material properties remains scarce (recent work has shown that the two-dimensional vacancy structures in Ga$_2$SeTe$_2$ can influence pressure-induced amorphization behavior \cite{AbdulJabbar:2014gt}) . Here, we attempted to correlate the effect of stoichiometric vacancy ordering on the electrical properties of Ga$_2$SeTe$_2$ via band gap and Hall effect measurements. First, we measured the band gap of an as-grown Ga$_2$SeTe$_2$ single crystal. The sample then underwent the 735 \degree C annealing treatment and was remeasured under identical conditions. Finally, a Ga$_2$SeTe$_2$ single crystal under the 435 \degree C annealing treatment was measured. The results are shown as absorption spectra in Figure 4, where the rise in absorption occurs at the band gap energy. The as-grown and 435 \degree C sample have a band gap of $\approx$1.27 eV. As the vacancies in Ga$_2$SeTe$_2$ become fully  ordered (via the 735 \degree C annealing treatment), the band gap of Ga$_2$SeTe$_2$ decreases to $\approx$1.22 eV. While this measured band gap change is slight, it appears to be an encouraging example of controlling intrinsic material parameters to engineer electronic structure (in contrast to extrinsic means like doping).

\begin{figure}
\includegraphics[width=\textwidth]{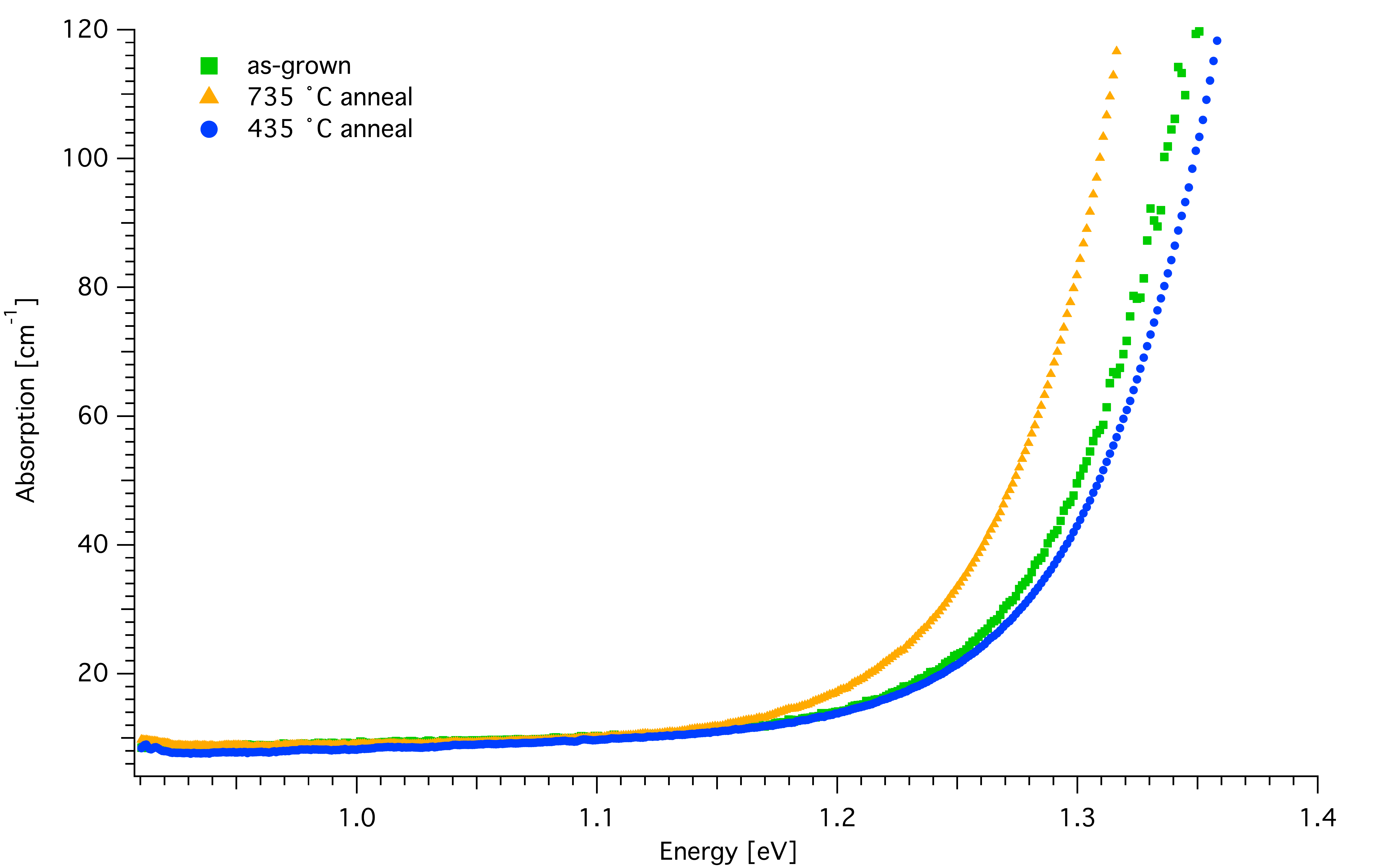}
\caption{Absorption edges of as-grown, 435 \degree C annealed, and 735 \degree C annealed Ga$_2$SeTe$_2$ single crystals.}
\label{fig4}
\end{figure}

Electrical properties of Ga$_2$SeTe$_2$ obtained from Hall effect experiments are shown Table I. The 435 \degree C sample had a resistivity of 3.1 M$\Omega$cm, the as-grown had $\approx$0.71 M$\Omega$cm, and the 735 \degree C sample $\approx$0.028 M$\Omega$cm. Similarly, the 735 \degree C showed the highest carrier mobility (30 cm$^2$/Vs, $n$-type), when compared to the as-grown and 435 \degree C samples. Based on these results, it appears that vacancy ordering improves the charge carrier transport properties, although other effects might be at play. For example, crystal quality can vary tremendously across a single ingot and the large concentrations of stoichiometric vacancies lead to a large number of free anions causing Ga$_2$SeTe$_2$ to be a highly compensated semiconductor, which explains why both $p$-type and $n$-type conductivities were observed.

\begin{table}
\caption{Electrical Properties of Ga$_2$SeTe$_2$  }
\centering
\scalebox{1}{
\begin{tabular}{p{2cm} p{2cm} p{2cm} p{2cm}}
\hline\hline
Thermal History   &   $\rho$ [M$\Omega$cm]   &   Type   &  $\mu$ [cm$^2$/Vs] \\ [0.5ex] 
\hline
As-grown & 0.71 & p & 15 \\
735 \degree C & 0.028 & n & 30 \\
435 \degree C & 3.1 & n & 0.8 \\[1ex]
\hline
\end{tabular}}
\label{table:nonlin}
\end{table}

\section{Conclusion}

The spatial distribution of stoichiometric vacancies appear to have measurable effects on the structure and properties of Ga$_2$SeTe$_2$. X-ray diffraction and electron microscopy showed the formation two-dimensional structures, where their spatial ordering is highly dependent on specimen thermal history. X-ray absorption data show that the vacancies in Ga$_2$SeTe$_2$ also give rise to local distortions within the zincblende lattice, though the overall cubic symmetry is preserved. Band gap and Hall effect measurements indicate that vacancy ordering may play a role in improving the electrical transport properties of Ga$_2$SeTe$_2$, providing an encouraging demonstration of engineering material electronic structure through intrinsic means.

\section{Acknowledgements}
The authors would like to acknowledge P. N. Valdivia for useful discussions and C. A. Ramsey for experimental assistance. N. M. A. acknowledges support from the Nuclear Nonproliferation International Safeguards Graduate Fellowship Program sponsored by the National Nuclear Security AdministrationÕs Next Generation Safeguards Initiative (NGSI). This work was supported by the US Department of Energy/NNSA/NA22 and carried out at the Lawrence Berkeley National Laboratory under Contract No. DE-AC02-05CH11231. Electron microscopy was performed at NCEM, which is supported by the Office of Science, Office of Basic Energy Sciences of the U.S. Department of Energy under Contract No. DE-AC02-05CH11231. A portion of this work was performed at the DuPont-Northwestern-Dow Collaborative Access Team (DND-CAT) located at Sector 5 of the Advanced Photon Source (APS). DND-CAT is supported by E.I. DuPont de Nemours and Co., The Dow Chemical Company and Northwestern University. Use of the APS, an Office of Science User Facility operated for the U.S. Department of Energy (DOE) Office of Science by Argonne National Laboratory, was supported by the U.S. DOE under Contract No. DE-AC02-06CH11357. Portions of this research were also carried out at the Stanford Synchrotron Radiation Lightsource, a Directorate of SLAC National Accelerator Laboratory and an Office of Science User Facility operated for the U.S. Department of Energy Office of Science by Stanford University.


\begin{thebibliography}{19}
\expandafter\ifx\csname natexlab\endcsname\relax\def\natexlab#1{#1}\fi
\expandafter\ifx\csname bibnamefont\endcsname\relax
  \def\bibnamefont#1{#1}\fi
\expandafter\ifx\csname bibfnamefont\endcsname\relax
  \def\bibfnamefont#1{#1}\fi
\expandafter\ifx\csname citenamefont\endcsname\relax
  \def\citenamefont#1{#1}\fi
\expandafter\ifx\csname url\endcsname\relax
  \def\url#1{\texttt{#1}}\fi
\expandafter\ifx\csname urlprefix\endcsname\relax\def\urlprefix{URL }\fi
\providecommand{\bibinfo}[2]{#2}
\providecommand{\eprint}[2][]{\url{#2}}

\bibitem[{\citenamefont{Kurosaki et~al.}(2008)\citenamefont{Kurosaki, Yamanaka,
  and Ishimaru}}]{Kurosaki:2008cj}
\bibinfo{author}{\bibfnamefont{K.}~\bibnamefont{Kurosaki}},
  \bibinfo{author}{\bibfnamefont{S.}~\bibnamefont{Yamanaka}}, \bibnamefont{and}
  \bibinfo{author}{\bibfnamefont{M.}~\bibnamefont{Ishimaru}},
  \bibinfo{journal}{Applied Physics Letters} \textbf{\bibinfo{volume}{93}}
  (\bibinfo{year}{2008}).

\bibitem[{\citenamefont{Kim et~al.}(2009)\citenamefont{Kim, Kurosaki, Ishimaru,
  Jung, Muta, and Yamanaka}}]{Kim:2009dz}
\bibinfo{author}{\bibfnamefont{C.-E.} \bibnamefont{Kim}},
  \bibinfo{author}{\bibfnamefont{K.}~\bibnamefont{Kurosaki}},
  \bibinfo{author}{\bibfnamefont{M.}~\bibnamefont{Ishimaru}},
  \bibinfo{author}{\bibfnamefont{D.-Y.} \bibnamefont{Jung}},
  \bibinfo{author}{\bibfnamefont{H.}~\bibnamefont{Muta}}, \bibnamefont{and}
  \bibinfo{author}{\bibfnamefont{S.}~\bibnamefont{Yamanaka}},
  \bibinfo{journal}{physica status solidi (RRL) - Rapid Research Letters}
  \textbf{\bibinfo{volume}{3}}, \bibinfo{pages}{221} (\bibinfo{year}{2009}).

\bibitem[{\citenamefont{Yamanaka et~al.}(2009)\citenamefont{Yamanaka, Ishimaru,
  Charoenphakdee, Matsumoto, and Kurosaki}}]{Yamanaka:2009ka}
\bibinfo{author}{\bibfnamefont{S.}~\bibnamefont{Yamanaka}},
  \bibinfo{author}{\bibfnamefont{M.}~\bibnamefont{Ishimaru}},
  \bibinfo{author}{\bibfnamefont{A.}~\bibnamefont{Charoenphakdee}},
  \bibinfo{author}{\bibfnamefont{H.}~\bibnamefont{Matsumoto}},
  \bibnamefont{and} \bibinfo{author}{\bibfnamefont{K.}~\bibnamefont{Kurosaki}},
  \bibinfo{journal}{Journal of Electronic Materials}
  \textbf{\bibinfo{volume}{38}}, \bibinfo{pages}{1392} (\bibinfo{year}{2009}).

\bibitem[{\citenamefont{Zhu et~al.}(2010)\citenamefont{Zhu, Yin, Xia, and
  Liu}}]{Zhu:2010jj}
\bibinfo{author}{\bibfnamefont{H.}~\bibnamefont{Zhu}},
  \bibinfo{author}{\bibfnamefont{J.}~\bibnamefont{Yin}},
  \bibinfo{author}{\bibfnamefont{Y.}~\bibnamefont{Xia}}, \bibnamefont{and}
  \bibinfo{author}{\bibfnamefont{Z.}~\bibnamefont{Liu}},
  \bibinfo{journal}{Applied Physics Letters} \textbf{\bibinfo{volume}{97}},
  \bibinfo{pages}{083504} (\bibinfo{year}{2010}).

\bibitem[{\citenamefont{Koshkin and Dmitriev}(1994)}]{Koshkin:1994}
\bibinfo{author}{\bibfnamefont{V.~M.} \bibnamefont{Koshkin}} \bibnamefont{and}
  \bibinfo{author}{\bibfnamefont{Y.~N.} \bibnamefont{Dmitriev}},
  \bibinfo{journal}{Chemistry Reviews} \textbf{\bibinfo{volume}{19}},
  \bibinfo{pages}{1} (\bibinfo{year}{1994}).

\bibitem[{\citenamefont{Koshkin et~al.}(1973)\citenamefont{Koshkin,
  Gal'chinetskii, Kulik, Minkov, and Ulmanis}}]{Koshkin:vp}
\bibinfo{author}{\bibfnamefont{V.~M.} \bibnamefont{Koshkin}},
  \bibinfo{author}{\bibfnamefont{L.~P.} \bibnamefont{Gal'chinetskii}},
  \bibinfo{author}{\bibfnamefont{V.~N.} \bibnamefont{Kulik}},
  \bibinfo{author}{\bibfnamefont{B.~I.} \bibnamefont{Minkov}},
  \bibnamefont{and} \bibinfo{author}{\bibfnamefont{U.~A.}
  \bibnamefont{Ulmanis}}, \bibinfo{journal}{Solid State Communications}
  \textbf{\bibinfo{volume}{13}}, \bibinfo{pages}{1} (\bibinfo{year}{1973}).

\bibitem[{\citenamefont{Koshkin et~al.}(1976)\citenamefont{Koshkin,
  Gal'chinetskii, Kulik, and Ulmanis}}]{Koshkin:1976wj}
\bibinfo{author}{\bibfnamefont{V.~M.} \bibnamefont{Koshkin}},
  \bibinfo{author}{\bibfnamefont{L.~P.} \bibnamefont{Gal'chinetskii}},
  \bibinfo{author}{\bibfnamefont{V.~N.} \bibnamefont{Kulik}}, \bibnamefont{and}
  \bibinfo{author}{\bibfnamefont{U.~A.} \bibnamefont{Ulmanis}},
  \bibinfo{journal}{Radiation Effects} \textbf{\bibinfo{volume}{29}},
  \bibinfo{pages}{1} (\bibinfo{year}{1976}).

\bibitem[{\citenamefont{Huang et~al.}(2014)\citenamefont{Huang, Abdul-Jabbar,
  and Wirth}}]{Huang:2014fo}
\bibinfo{author}{\bibfnamefont{G.-Y.} \bibnamefont{Huang}},
  \bibinfo{author}{\bibfnamefont{N.~M.} \bibnamefont{Abdul-Jabbar}},
  \bibnamefont{and} \bibinfo{author}{\bibfnamefont{B.~D.} \bibnamefont{Wirth}},
  \bibinfo{journal}{Acta Materialia} \textbf{\bibinfo{volume}{71}},
  \bibinfo{pages}{349} (\bibinfo{year}{2014}).

\bibitem[{\citenamefont{Warren}(1974)}]{Warren:1974ux}
\bibinfo{author}{\bibfnamefont{W.~W.} \bibnamefont{Warren}},
  \bibinfo{journal}{Journal of Physics and Chemistry of Solids}
  \textbf{\bibinfo{volume}{35}}, \bibinfo{pages}{1153} (\bibinfo{year}{1974}).

\bibitem[{\citenamefont{Abdul-Jabbar et~al.}(2012)\citenamefont{Abdul-Jabbar,
  Bourret-Courchesne, and Wirth}}]{AbdulJabbar:2012bma}
\bibinfo{author}{\bibfnamefont{N.}~\bibnamefont{Abdul-Jabbar}},
  \bibinfo{author}{\bibfnamefont{E.~D.} \bibnamefont{Bourret-Courchesne}},
  \bibnamefont{and} \bibinfo{author}{\bibfnamefont{B.~D.} \bibnamefont{Wirth}},
  \bibinfo{journal}{Journal of Crystal Growth} \textbf{\bibinfo{volume}{352}},
  \bibinfo{pages}{31} (\bibinfo{year}{2012}).

\bibitem[{\citenamefont{Kim et~al.}(2011)\citenamefont{Kim, Kurosaki, Ishimaru,
  Muta, and Yamanaka}}]{Kim:2011kb}
\bibinfo{author}{\bibfnamefont{C.-E.} \bibnamefont{Kim}},
  \bibinfo{author}{\bibfnamefont{K.}~\bibnamefont{Kurosaki}},
  \bibinfo{author}{\bibfnamefont{M.}~\bibnamefont{Ishimaru}},
  \bibinfo{author}{\bibfnamefont{H.}~\bibnamefont{Muta}}, \bibnamefont{and}
  \bibinfo{author}{\bibfnamefont{S.}~\bibnamefont{Yamanaka}},
  \bibinfo{journal}{Journal of Electronic Materials}
  \textbf{\bibinfo{volume}{40}}, \bibinfo{pages}{999} (\bibinfo{year}{2011}).

\bibitem[{\citenamefont{Karapetrova et~al.}(2011)\citenamefont{Karapetrova,
  Ice, Tischler, Hong, and Zschack}}]{Karapetrova:2011tb}
\bibinfo{author}{\bibfnamefont{E.}~\bibnamefont{Karapetrova}},
  \bibinfo{author}{\bibfnamefont{G.}~\bibnamefont{Ice}},
  \bibinfo{author}{\bibfnamefont{J.}~\bibnamefont{Tischler}},
  \bibinfo{author}{\bibfnamefont{H.}~\bibnamefont{Hong}}, \bibnamefont{and}
  \bibinfo{author}{\bibfnamefont{P.}~\bibnamefont{Zschack}},
  \bibinfo{journal}{Nuclear Instruments and Methods in Physics Research A}
  \textbf{\bibinfo{volume}{649}}, \bibinfo{pages}{52} (\bibinfo{year}{2011}).

\bibitem[{\citenamefont{Webb}(2005)}]{Webb:2005jv}
\bibinfo{author}{\bibfnamefont{S.~M.} \bibnamefont{Webb}},
  \bibinfo{journal}{Physica Scripta} \textbf{\bibinfo{volume}{2005}},
  \bibinfo{pages}{1011} (\bibinfo{year}{2005}).

\bibitem[{\citenamefont{Abdul-Jabbar
  et~al.}(2014{\natexlab{a}})\citenamefont{Abdul-Jabbar, Ercius, Gronsky,
  Bourret-Courchesne, and Wirth}}]{AbdulJabbar:2014fl}
\bibinfo{author}{\bibfnamefont{N.~M.} \bibnamefont{Abdul-Jabbar}},
  \bibinfo{author}{\bibfnamefont{P.}~\bibnamefont{Ercius}},
  \bibinfo{author}{\bibfnamefont{R.}~\bibnamefont{Gronsky}},
  \bibinfo{author}{\bibfnamefont{E.~D.} \bibnamefont{Bourret-Courchesne}},
  \bibnamefont{and} \bibinfo{author}{\bibfnamefont{B.~D.} \bibnamefont{Wirth}},
  \bibinfo{journal}{Applied Physics Letters} \textbf{\bibinfo{volume}{104}},
  \bibinfo{pages}{051904} (\bibinfo{year}{2014}{\natexlab{a}}).

\bibitem[{\citenamefont{Otaki et~al.}(2009{\natexlab{a}})\citenamefont{Otaki,
  Yanadori, Seki, Tadano, and Kashida}}]{Otaki:2009if}
\bibinfo{author}{\bibfnamefont{Y.}~\bibnamefont{Otaki}},
  \bibinfo{author}{\bibfnamefont{Y.}~\bibnamefont{Yanadori}},
  \bibinfo{author}{\bibfnamefont{Y.}~\bibnamefont{Seki}},
  \bibinfo{author}{\bibfnamefont{M.}~\bibnamefont{Tadano}}, \bibnamefont{and}
  \bibinfo{author}{\bibfnamefont{S.}~\bibnamefont{Kashida}},
  \bibinfo{journal}{Journal of Solid State Chemistry}
  \textbf{\bibinfo{volume}{182}}, \bibinfo{pages}{1556}
  (\bibinfo{year}{2009}{\natexlab{a}}).

\bibitem[{\citenamefont{Otaki et~al.}(2009{\natexlab{b}})\citenamefont{Otaki,
  Yanadori, Seki, Yamamoto, and Kashida}}]{Otaki:2009cg}
\bibinfo{author}{\bibfnamefont{Y.}~\bibnamefont{Otaki}},
  \bibinfo{author}{\bibfnamefont{Y.}~\bibnamefont{Yanadori}},
  \bibinfo{author}{\bibfnamefont{Y.}~\bibnamefont{Seki}},
  \bibinfo{author}{\bibfnamefont{K.}~\bibnamefont{Yamamoto}}, \bibnamefont{and}
  \bibinfo{author}{\bibfnamefont{S.}~\bibnamefont{Kashida}},
  \bibinfo{journal}{Acta Materialia} \textbf{\bibinfo{volume}{57}},
  \bibinfo{pages}{1392} (\bibinfo{year}{2009}{\natexlab{b}}).

\bibitem[{\citenamefont{Kashida et~al.}(2009)\citenamefont{Kashida, Otaki,
  Yanadori, Seki, and Tadano}}]{Kashida:2009bc}
\bibinfo{author}{\bibfnamefont{S.}~\bibnamefont{Kashida}},
  \bibinfo{author}{\bibfnamefont{Y.}~\bibnamefont{Otaki}},
  \bibinfo{author}{\bibfnamefont{Y.}~\bibnamefont{Yanadori}},
  \bibinfo{author}{\bibfnamefont{Y.}~\bibnamefont{Seki}}, \bibnamefont{and}
  \bibinfo{author}{\bibfnamefont{M.}~\bibnamefont{Tadano}},
  \bibinfo{journal}{physica status solidi c} \textbf{\bibinfo{volume}{6}},
  \bibinfo{pages}{1162} (\bibinfo{year}{2009}).

\bibitem[{\citenamefont{Als-Nielsen and McMorrow}(2011)}]{AlsNielsen:2011vn}
\bibinfo{author}{\bibfnamefont{J.}~\bibnamefont{Als-Nielsen}} \bibnamefont{and}
  \bibinfo{author}{\bibfnamefont{D.}~\bibnamefont{McMorrow}},
  \emph{\bibinfo{title}{{Elements of Modern X-ray Physics}}}
  (\bibinfo{publisher}{Wiley}, \bibinfo{year}{2011}), \bibinfo{edition}{2nd}
  ed.

\bibitem[{\citenamefont{Abdul-Jabbar
  et~al.}(2014{\natexlab{b}})\citenamefont{Abdul-Jabbar, Kalkan, Huang,
  Gronsky, MacDowell, Bourret-Courchesne, and Wirth}}]{AbdulJabbar:2014gt}
\bibinfo{author}{\bibfnamefont{N.~M.} \bibnamefont{Abdul-Jabbar}},
  \bibinfo{author}{\bibfnamefont{B.}~\bibnamefont{Kalkan}},
  \bibinfo{author}{\bibfnamefont{G.~Y.} \bibnamefont{Huang}},
  \bibinfo{author}{\bibfnamefont{R.}~\bibnamefont{Gronsky}},
  \bibinfo{author}{\bibfnamefont{A.~A.} \bibnamefont{MacDowell}},
  \bibinfo{author}{\bibfnamefont{E.~D.} \bibnamefont{Bourret-Courchesne}},
  \bibnamefont{and} \bibinfo{author}{\bibfnamefont{B.~D.} \bibnamefont{Wirth}},
  \bibinfo{journal}{Applied Physics Letters} \textbf{\bibinfo{volume}{105}},
  \bibinfo{pages}{051908} (\bibinfo{year}{2014}{\natexlab{b}}).

\end{thebibliography}

\end{document}